\documentclass[11pt, reqno]{amsart}
\usepackage{hyperref}
\usepackage{amsmath, amsthm, amscd, amsfonts, amssymb, graphicx, color}

\begin{document}
\title[Integrals of Higher Binary Options and Defaultable Bonds...]
{Integrals of Higher Binary Options and Defaultable Bonds with Discrete Default Information}

\author[Hyong-Chol O, Dong-Hyok Kim, Jong-Jun Jo and Song-Hun Ri]{Hyong-Chol O, Dong-Hyok Kim, Jong-Jun Jo and Song-Hun Ri}

\address{Hyong-Chol O, {\it Corresponding author}  \newline
Faculty of Mathematics,  Kim Il Sung University,  Pyongyang, D. P. R. Korea}
\email{ohyongchol@yahoo.com}

\address{Dong-Hyok Kim \newline
Faculty of Mathematics,  Kim Il Sung University,  Pyongyang, D. P. R. Korea}

\address{Jong-Jun Jo  \newline
Faculty of Mathematics,  Kim Il Sung University,  Pyongyang, D. P. R. Korea}

\address{Song-Hun Ri  \newline
Faculty of Mathematics, Wonsan Educational University, Wonsan, D. P. R. Korea}

\thanks{First version submitted: on May 30, 2013; last revised: on 21 Oct. 2013.}

\setcounter{page}{1}

\maketitle
\begin{abstract}
In this article, we study the problem of pricing defaultable bond with discrete default intensity and barrier under constant risk free short rate using higher order binary options and their integrals. In our credit risk model, the risk free short rate is a constant and the default event occurs in an expected manner when the firm value reaches a given default barrier at predetermined discrete announcing dates or in an unexpected manner at the first jump time of a Poisson process with given default intensity given by a step function of time variable, respectively. We consider both endogenous and exogenous default recovery. Our pricing problem is derived to a solving problem of inhomogeneous or homogeneous Black-Scholes PDEs with different coefficients and terminal value of binary type in every subinterval between the two adjacent announcing dates. In order to deal with the difference of coefficients in subintervals we use a relation between prices of higher order binaries with different coefficients. In our model, due to the inhomogenous term related to endogenous recovery, our pricing formulae are represented by not only the prices of higher binary options but also the integrals of them. So we consider a special binary option called integral of i-th binary or nothing and then we obtain the pricing formulae of our defaultable corporate bond by using the pricing formulae of higher binary options and integrals of them.
\end{abstract}

\section{Introduction}
The study on defaultable corporate bond and credit risk is now one of the most promising areas of cutting edge in financial mathematics. There are two main approaches to modeling credit risk and pricing defaultable corporate bonds; one is the {\it structural approach} and the other one is the {\it reduced form approach}. In the structural method, we think that the default event occurs when the firm value is not enough to repay debt, that is, the firm value reaches a certain lower threshold ({\it default barrier}) from the above. Such a default can be expected and thus we call it {\it expected default}. In the reduced-form approach, the default is treated as an unpredictable event governed by a default intensity process. In this case, the default event can occur without any correlation with the firm value and such a default is called {\it unexpected default}. In the reduced-form approach, if the default probability in time interval  $[t,~t+\Delta t]$ is , then $\lambda \Delta t$ is $\lambda$ called {\it default intensity} \cite{ohc, OW, OWR, wil}.

The two approaches have got their own advantages and shortcomings (\cite{BaP, OW}) and therefore the use of unified models of structural approach and reduced-form approach is a trend. (See \cite{BaP, BiB, CaE1, CaE2, ohc, OW, OWR, rea}.) Cathcart et al \cite{CaE1} studied a pricing of corporate bonds in the case when the default intensity is a linear function of the interest rate and gave semi-analytical pricing formulae. Cathcart et al \cite{CaE2} studied a valuation model in the case when the default intensity (hazard rate) is a linear function of the state variable and the interest rate. Realdon \cite{rea} studied a pricing of corporate bonds in the case with constant default intensity and gave pricing formulae of the bond using PDE method. Some authors studied the pricing model of defaultable bonds in which the default intensity is given as a stochastic process \cite{BaP, BiB, OW}.  In \cite{OW}, the authors provided analytical pricing formula of corporate defaultable bond with both expected and unexpected default in the case when stochastic default intensity follows Wilmott model where drift and volatility are linear of state variables \cite{wil}. Bi et all \cite{BiB} got the similar result with \cite{OW} in the case when stochastic default intensity follows CIR-like model. Ballestra et al \cite{BaP} proposed a model to price defaultable bonds where default intensity follows Vasicek-like model or CIR-like model coupled with the process of the firm's asset value and provided a closed-form approximate solution to their model. In \cite{BaP, BiB, CaE1, CaE2, OW, rea} expected default barrier is given in the whole lifetime of the bond. 

On the other hand, in \cite{ohc, OWR} the author studied the pricing problem for defaultable corporate bond under the assumption that we only know the firm value and the default barrier at 2 fixed discrete announcing dates, we don't know about any information of the firm value in another time and the default intensity between the adjoined two announcing dates is a constant determined by its announced firm value at the former announcing date. The computational error in \cite{OWR} is corrected in \cite{ohc}. The approach of \cite{ohc, OWR} is a kind of study of defaultable bond under {\it insufficient information} about the firm and it is interesting to note that Agliardi et al \cite{AA} studied bond pricing problem under {\it imprecise information} with the technique of fuzzy mathematics. The approach of \cite{ohc, OWR} can be seen as a {\it unified model} of structural model and reduced form model. Agliardi \cite{agl} studied a {\it structural model} for defaultable bond with several (discrete) coupon dates where the default can occur only when the firm value is not large enough to pay its debt and coupon in those {\it discrete coupon dates}.   

Speaking on default recovery, most of authors including \cite{AA, BaP, BiB, CaE1, ohc, OW, OWR} have studied the case of {\it exogenous} default recovery which is independent on firm value whereas \cite{agl} have studied the case of {\it endogenous} recovery which is related to firm value, and \cite{rea} studied both cases of exogenous and endogenous recovery. 

Here we study the problem of pricing defaultable bond with discrete default intensity and barrier under constant risk free short rate using higher order binary options and their integrals. In our credit risk model, the default event occurs in an expected manner when the firm value reaches a certain lower threshold - the default barrier at predetermined discrete announcing dates or in an unexpected manner at the first jump time of a Poisson process with given default intensity given by a step function of time variable, respectively. We consider both {\it endogenous} and {\it exogenous} default recovery. Our pricing problem is derived to a solving problem of {\it inhomogeneous} or {\it homogeneous Black-Scholes PDEs} with {\it different coefficients} and terminal value of binary type in every subinterval between the two adjacent announcing dates. In order to deal with the difference of coefficients in subintervals we use a {\it relation} between prices of higher order binaries with different coefficients. In our model, due to the inhomogenous term related to endogenous recovery, our pricing formulae are represented by not only the prices of higher binary options but also the integrals of them. So we consider a {\it special binary} option called {\it integral of i-th binary} or {\it nothing} and then we obtain the pricing formulae of our defaultable corporate bond by using the pricing formulae of higher binary options and integrals of them.

Our approach to model credit risk is similar with the one of \cite{OWR, ohc}. One of the different points of our model from \cite{ohc} is that we here consider arbitrary number of announcing dates but \cite{ohc} consider only 2 announcing dates. Another different point from \cite{ohc} is that we use constant risk free rate, the purpose of which is to show the applicability of higher order binaries to the pricing of defaultable bonds in the simplest way. Unlikely in \cite{ohc} we here consider discrete default intensity independent on firm value and it can be seen incompatible with reality but we think our analytical pricing formulae can help the further study on the more realistic situation with discrete default intensity dependent on firm value.

The remainder of the article is organized as follows. In section 2 we give some preliminary knowledge on prices of higher order binary options and their integral on the last expiry date. In section 3 we set our problem for corporate defaultable bonds, provide the pricing formulae in both cases of endogenous and exogenous default recovery and analyze the credit spread. In section 4 we derive the pricing formulae using and higher order binary options and their integral. 

\section{Preliminaries and Notes on Binary Options and their Integrals}
First, we introduce the concept of higher order bond and asset binaries with risk free rate $r$, dividend rate $q$ and volatility $\sigma$ and their pricing formulae \cite{buc, OK1, OK2}.
\begin{equation}\label{eq2-1}
\frac{\partial V}{\partial t}+\frac{\sigma^{2}}{2}x^2\frac{\partial^2 V}{\partial x}+(r-q)x\frac{\partial V}{\partial x}-rV=0,\quad 0\leq t<T,~0<x<\infty,       
\end{equation}
\begin{equation}\label{eq2-2}
V(x,~T)=x\cdot 1(sx>s\xi),
\end{equation}
\begin{equation}\label{eq2-3}
V(x,~T)=1(sx>s\xi).
\end{equation}
The solution to the problem \eqref{eq2-1} and \eqref{eq2-2} is called the {\it asset-or-nothing} binaries (or {\it asset} binaries) and denoted by $A^s_\xi (x,t;T)$. The solution to the problem \eqref{eq2-1} and \eqref{eq2-3} is called the {\it cash-or-nothing} binaries (or {\it bond} binaries) and denoted by $B^s_\xi (x,t;T)$. Asset binary and bond binary are called the {\it first order binary} options. If necessary, we will denote by $A^s_\xi (x,t;T; r,q,\sigma)$ or $B^s_\xi (x,t;T; r,q,\sigma)$ where the coefficients $r$, $q$ and $\sigma$ of Black-Scholes equation \eqref{eq2-1} are explicitly included in the notation.

Let assume that $0<T_0<T_1<\cdots<T_{n-1}$ and the $(n-1)$-th order (asset or bond) binary options $A^{s_1\cdots s_{n-1}}_{\xi_1\cdots \xi_{n-1}}(x,t;T_1,\cdots,T_{n-1})$ and $B^{s_1\cdots s_{n-1}}_{\xi_1\cdots \xi_{n-1}}(x,t;T_1,\cdots$ $,T_{n-1})$ are already defined. Let 
\begin{equation}\label{eq2-4}
V(x,~T_0)=A^{s_1\cdots s_{n-1}}_{\xi_1\cdots \xi_{n-1}}(x,T_0;T_1,\cdots,T_{n-1})\cdot 1(s_0x>s_0\xi_0),
\end{equation}
\begin{equation}\label{eq2-5}
V(x,~T_0)=B^{s_1\cdots s_{n-1}}_{\xi_1\cdots \xi_{n-1}}(x,T_0;T_1,\cdots,T_{n-1})\cdot 1(s_0x>s_0\xi_0).
\end{equation}
The solution to the problem \eqref{eq2-1} and \eqref{eq2-4} is called the {\it n-th order asset binaries} and denoted by $A^{s_0s_1\cdots s_{n-1}}_{\xi_0\xi_1\cdots \xi_{n-1}}(x,t;T_0,T_1,\cdots,T_{n-1})$. The solution to the problem \eqref{eq2-1} and \eqref{eq2-5} is called the {\it n-th order bond binaries} and denoted by $B^{s_0s_1\cdots s_{n-1}}_{\xi_0\xi_1\cdots \xi_{n-1}}(x,t;T_0,T_1,\cdots,T_{n-1})$.\\

{\bf Lemma 1}. (The pricing formulae of higher order binary options) \cite{buc, OK1, OK2} {\it The prices of higher order bond and asset binaries with risk free rate $r$, dividend rate $q$ and volatility $\sigma$ are as follows}.
\begin{align}\label{eq2-6}
\nonumber &A^s_\xi (x,t;T; r,q,\sigma)=xe^{-q(T-t)}N(sd^+),\\
&B^s_\xi (x,t;T; r,q,\sigma)=e^{-r(T-t)}N_1(sd^-), s=+~\mathtt{or}~-.
\end{align}
{\it Here}
\begin{align*}
&N_1(x)=(\sqrt{2\pi})^{-1}\int_{-\infty}^{x}exp(-y^2/2)dy,\\
&d^{\pm}=(\sigma\sqrt{T-t})^{-1}[\ln(x/K)+(r-q\pm\sigma^2/2)(T-t)].
\end{align*}
\begin{align}\label{eq2-7}
\nonumber &A^{s_1~s_2}_{K_1K_2}(x,t;T_1,T_2; r,q,\sigma)=xe^{-q(T_2-t)}N_2(s_1d^+_1,s_2d^+_2;s_1s_2\rho),\\
&B^{s_1~s_2}_{K_1K_2}(x,t;T_1,T_2; r,q,\sigma)=e^{-r(T_2-t)}N_2(s_1d^-_1,s_2d^-_2;s_1s_2\rho), s_1,s_2=+~\mathtt{or}~-.
\end{align}
{\it Here}
\begin{align*}
&N_2(a,b;\rho)=\int_{-\infty}^{a}\int_{-\infty}^{b}(2\pi\sqrt{1-\rho^2})^{-1}e^{-\frac{y^2-2\rho yz+z^2}{2(1-\rho^2)}}dydz,\\
&d^{\pm}_i=(\sigma\sqrt{T_i-t})^{-1}[\ln(x/K_i)+(r-q\pm\sigma^2/2)(T_i-t)],i=1,2,\\
&\rho=\sqrt{(T_1-t)/(T_2-t)}.
\end{align*}
{\it If $m>2$ and $s_i=+$ or $-$, $i=1,\cdots,m$, then we have}
\begin{align}\label{eq2-8}
\nonumber &A^{s_1\cdots s_m}_{K_1\cdots K_m}(x,t;T_1,\cdots,T_m; r,q,\sigma)=xe^{-q(T_m-t)}N_m(s_1d^+_1,\cdots,s_md^+_m;A_{s_1\cdots s_m}),\\
&B^{s_1\cdots s_m}_{K_1\cdots K_m}(x,t;T_1,\cdots,T_m; r,q,\sigma)=e^{-r(T_m-t)}N_m(s_1d^-_1,\cdots,s_md^-_m;A_{s_1\cdots s_m}).
\end{align}
{\it Here}
\begin{align}\label{eq2-9}
\nonumber &N_m(a_1,\cdots,a_m;A)=\int_{-\infty}^{a_1}\cdots\int_{-\infty}^{a_m}(\sqrt{2\pi})^{-m}\sqrt{\det A}\exp\left(-\frac{1}{2}y^\mathsf{T} Ay\right)dy,\\
\nonumber &d^{\pm}_i=(\sigma\sqrt{T_i-t})^{-1}[\ln(x/K_i)+(r-q\pm\sigma^2/2)(T_i-t)],i=1,\cdots,m,\\
&A_{s_1\cdots s_m}=(s_is_ja_{ij})_{i,j=1}^{m},~y^\mathsf{T}=(y_1,\cdots,y_m),
\end{align}
and the matrix $(a_{i,j})_{i,j=1}^{m}$ is given as follows:
\begin{align}\label{eq2-10}
\nonumber &a_{11}=(T_2-t)/(T_2-T_1),~~~a_{mm}=(T_m-t)/(T_m-T_{m-1}),\\
\nonumber &a_{ii}=(T_i-t)/(T_i-T_{i-1})+(T_i-t)/(T_{i+1}-T_{i}),~2\leq i\leq m-1,\\
\nonumber &a_{i,i+1}=a_{i+1,i}=-\sqrt{(T_i-t)(T_{i+1}-t)}/(T_{i+1}-T_{i}),~1\leq i\leq m-1,\\
&a_{ij}=0~\mathtt{~for~another~}~ i,j =1,\cdots,m.
\end{align}
{\it Note} that $N_2(a,b;\rho)$ is the {\it cumulative distribution function of bivariate normal distribution} with a mean vector $[0, 0]$ and a {\it covariance matrix} $[1, \rho; \rho, 1]$  (symbols in {\bf Mat lab}), and $N_m(a_1,\cdots,a_m;A)$ is the {\it cumulative distribution function of m-variate normal distribution} with {\bf zero} {\it mean vector} and a {\it covariance matrix} $A^{-1}=(r_{ij})_{i,j=1}^{m}$  where $r_{ij}=\sqrt{(T_i-t)/(T_j-t)},~r_{ji}=r_{ij}, i\leq j$. Such special functions can easily be calculated by standard functions supplied in software for mathematical calculation (for example, {\bf Mat lab}). Note that $(A_{s_1\cdots s_m})^{-1}=(s_is_jr_{ij})_{i,j=1}^m$.

Second, we consider a relation between prices of higher order binaries with different risk free rates and dividend rates. From the formulae \eqref{eq2-6}, \eqref{eq2-7} and \eqref{eq2-8}, we can easily know that the {\it following relations} between prices of {\it higher order binaries} with {\it different} risk free rates and dividend rates hold:
\begin{align}\label{eq2-11}
\nonumber F^{s_1\cdots s_m}_{K_1\cdots K_m}&(x,t;T_1,\cdots,T_m; r_1,r_1+b,\sigma)=\\
&=e^{-(r_1-r_2)(T_m-t)}F^{s_1\cdots s_m}_{K_1\cdots K_m}(x,t;T_1,\cdots,T_m; r_2,r_2+b,\sigma).
\end{align}
Here $F=A$ or $F=B$.

Next, we will discuss {\it integrals} of the prices of higher order binary options on {\it the last expiry date variable}. Let consider \eqref{eq2-1} with the following two terminal conditions:
\begin{equation}\label{eq2-12}
V(x,~T)=f(x,~\tau),
\end{equation}
\begin{equation}\label{eq2-13}
V(x,~T)=F(x):=\int_C^Df(x,~\tau)d\tau.
\end{equation}
   
{\bf Lemma 2}. {\it Assume that there exist non negative constants $M$ and $\alpha$ such that     $|f(x,\tau)|\leq M\cdot x^{\alpha\ln x},~x>0$ and $f(x,\tau)$ is a continuous function of $\tau\in [C,~D]$. Then the solution $V_F(x,t)$ to the problem \eqref{eq2-1} and \eqref{eq2-13} is given by the integral of the solution $V_f(x,t;\tau)$ to the problem \eqref{eq2-1} and \eqref{eq2-12}}:
\begin{equation}\label{eq2-14}
V_F(x,~t)=\int_C^DV_f(x,~t;~\tau)d\tau.
\end{equation}

{\bf Proof}: If we use the proposition 1 at page 249 in \cite{OK1} and the continuity of $f$ on $\tau$, we can easily get \eqref{eq2-14}.(QED)   \\

Now let consider a {\it special} binary option called {\it integral of i-th binary or nothing}.\\

{\bf Corollary.} Let $g(\tau)$ be a continuous function of $\tau\in[T_{i-1},T]$ and
\begin{equation}\label{eq2-15}
V(x,~T_0)=1(s_0x>s_0K_0)\int_{T_{i-1}}^Tg(\tau)F^{s_1\cdots s_{i-1}~s_i}_{K_1\cdots K_{i-1}K_i}(x,T_0;T_1,\cdots,T_{i-1},\tau)d\tau.
\end{equation}
{\it Then the solution of \eqref{eq2-1} and \eqref{eq2-15} is given as follows}:
\begin{equation}\label{eq2-16}
V(x,~t)=\int_{T_{i-1}}^Tg(\tau)F^{s_0~s_1\cdots s_{i-1}~s_i}_{K_0K_1\cdots K_{i-1}K_i}(x,t;T_0,T_1,\cdots,T_{i-1},\tau)d\tau,~t<T_0.
\end{equation}
{\it Here $F=A$ or $F=B$}.

   {\bf Proof}: We will prove only for bond binary in the case when $i =1$. The proofs for other cases are the same. By the proposition 1 at page 249 in \cite{OK1}, the solution to \eqref{eq2-1} with 
\begin{equation*}
V(x,~T_0)=1(s_0x>s_0K_0)\int_{T_{0}}^Tg(\tau)B^{s_1}_{K_1}(x,T_0;\tau)d\tau
\end{equation*}
is given as follows:
\begin{align*}
&U(x,~t)=\\
=&\frac{e^{-r(T_0-t)}}{\sigma \sqrt{2 \pi (T_0-t)}}\int_{0}^{\infty} \frac{1}{z}e^{-\frac{[\ln \frac{x}{z}+(r-q- \frac{\sigma^{2}}{2})(T_0-t)]^2}{2 \sigma^2 (T_0-t)}}1(s_0z>s_0K_0)\int_{T_{0}}^Tg(\tau)B^{s_1}_{K_1}(z,T_0;\tau)d\tau dz\\
=&\int_{T_{0}}^Tg(\tau)\frac{e^{-r(T_0-t)}}{\sigma \sqrt{2 \pi (T_0-t)}}\int_{0}^{\infty} \frac{1}{z}e^{-\frac{[\ln \frac{x}{z}+(r-q- \frac{\sigma^{2}}{2})(T_0-t)]^2}{2 \sigma^2 (T_0-t)}}B^{s_1}_{K_1}(z,T_0;\tau)1(s_0z>s_0K_0)dzd\tau\\
=&\int_{T_{0}}^Tg(\tau)B^{s_0~s_1}_{K_0K_1}(x,t;T_0,\tau)d\tau. ~~~\mathtt{(QED)}
\end{align*}

\section{The Problem of Defaultable Bonds and The Pricing Formulae}
\subsection{The Problem with Endogenous Recovery}
Let Assume the followings: 

1) Short rate $r$ is a constant.

2) $0=t_0<t_1<\cdots<t_{N_1}<t_N=T$ are announcing dates and $T$ is the maturity of our corporate bond with face value 1 (unit of currency). For every $i=0,\cdots,N-1$  the unexpected default probability in the interval $[t,t+dt]\cap(t_i,t_{i+1})$ is $\lambda_idt$. Here the {\it default intensity} $\lambda_i$ is a constant.

3) The firm value $V(t)$ follows a geometric Brownian motion 
\begin{equation}\label{eq3-1}
dV(t)=(r-b)dtV(t)dt+s_V\cdot V(t)dW(t)
\end{equation}
under the risk neutral martingale measure and the firm continuously pays out dividend in rate $b$ (constant) for a unit of firm value. Like in \cite{OW}, the firm value $V_t$ is assumed to consist of $m$ shares of stock $S$ and $n$ sheets of corporate bonds $C_t$:
\begin{equation}\label{eq3-2}
V_t=mS_t+nC_t.
\end{equation}

4) The expected default barrier is only given at time $t_i$ and the expected default event occurs when
\begin{equation}\label{eq3-3}
V(t_i)\leq K_ie^{-r(T-t_i)},~i=1,\cdots,N.
\end{equation}
Here $K_i$ is a constant reflecting the quantity of debt and $e^{-r(T-t_i)}$ is default free zero coupon bond price.

5) The default recovery $R_d$ is given as the form of endogenous face value
\begin{equation}\label{eq3-4}
R_d=\min\{e^{-r(T-t)}, R\cdot V/n\}.
\end{equation}
Here {\it recovery rate} $0\leq R\leq 1$ is a constant.

   6) In the subinterval $(t_i,t_{i+1})$, the price of our corporate bond is given by a sufficiently smooth function $C_i(V,~t)$ ($i=0,\cdots,N-1$).

{\bf Problem}: {\it Find the representation of the price function $C_i(V,~t)$ ($i=0,\cdots,N-1$) under the above assumptions}. \\

{\bf The Pricing Model.}
According to \cite{wil}, under the above assumptions the price $C$ of defaultable bond with a constant default intensity $\lambda$ and default recovery $R_d$ satisfies the following PDE:
\begin{equation*}
\frac{\partial C}{\partial t}+\frac{s_V^{2}}{2}V^2\frac{\partial^2 C}{\partial V^2}+(r-b)V\frac{\partial C}{\partial V}-(r+\lambda)C+\lambda R_d=0.          
\end{equation*}
Therefore if we let $C_N(V,  t)\equiv 1$ , then the {\it price model} of our bond is given as follows:
\begin{align}\label{eq3-5}
\nonumber &\frac{\partial C_i}{\partial t}+\frac{s_V^{2}}{2}V^2\frac{\partial^2 C_i}{\partial V^2}+(r-b)V\frac{\partial C_i}{\partial V}-(r+\lambda_i)C_i+\lambda_i\min\{e^{-r(T-t)}, \frac{RV}{n}\}=0,~t_i<t<t_{i+1}, \\
&C_i(t_{i+1})=C_{i+1}(t_{i+1})1(V>K_{i+1}e^{-r(T-t_{i+1})})+\min\{e^{-r(T-t_{i+1})}, \frac{RV}{n}\}1(V\leq K_{i+1}e^{-r(T-t_{i+1})}).
\end{align}
Here $i=0,1,\cdots,N-1$.\\

{\bf The Pricing Formulae.}
Under the assumptions 1)-- 6), we have the following pricing formulae:\\

{\bf Theorem 1}. (endogenous recovery) i) {\it Assume that $K_i\leq n/R, i=1,\cdots,N$. Under the assumptions 1)--6), the price of our bond, that is, the solution of \eqref{eq3-5} is represented as follows}: 
\begin{equation}\label{eq3-6}
C_i(V, t)=e^{-r(T-t)}u_i(V/e^{-r(T-t)}, t), t_i\leq t<t_{i+1},~i=0,\cdots,N-1.
\end{equation}
{\it Here}
\begin{align}\label{eq3-7}
\nonumber &u_i(x,t)=e^{-\lambda_i(t_{i+1}-t)}\left\{e^{-\sum_{k=i+1}^{N-1}\lambda_k(t_{k+1}-t_k)}B^{+\quad\cdots~+}_{K_{i+1}\cdots K_N}(x,t;t_{i+1},\cdots,t_N)+\right. \\
\nonumber &\quad+\frac{R}{n}\sum_{m=i}^{N-1}e^{-\sum_{k=i+1}^{m}\lambda_k(t_{k+1}-t_k)}A^{+\quad\cdots~+\quad-}_{K_{i+1}\cdots K_mK_{m+1}}(x,t;t_{i+1},\cdots,t_m,t_{m+1}) \\
\nonumber&\quad+\sum_{m=i+1}^{N-1}\lambda_me^{-\sum_{k=i+1}^{m-1}\lambda_k(t_{k+1}-t_k)}\int_{t_m}^{t_{m+1}}e^{-\lambda_m(\tau-t_m)}\left[B^{+\quad\cdots~+~+}_{K_{i+1}\cdots K_m\frac{n}{R}}(x,t;t_{i+1},\cdots,t_m,\tau) \right.\\
\nonumber&\quad\quad\quad\quad\quad\quad\quad\quad\quad\quad+\left.\frac{R}{n}\left.A^{+\quad\cdots~+~~-}_{K_{i+1}\cdots K_m\frac{n}{R}}(x,t;t_{i+1},\cdots,t_m,\tau)\right]d\tau\right\}\\
&+\lambda_i\int_{t}^{t_{i+1}}e^{-\lambda_i(\tau-t)}\left[B^{+}_{\frac{n}{R}}(x,t;\tau; 0,b,s_V)+\frac{R}{n}A^{-}_{\frac{n}{R}}(x,t;\tau; 0,b,s_V)\right]d\tau.
\end{align}
ii) {\it Assume that $K_i>n/R, i=1,\cdots,N$. Under the assumptions 1)--6), the price of our bond, that is, the solution of \eqref{eq3-5} is represented by \eqref{eq3-6} with the following} $u_i(x,t)$:  
\begin{align}\label{eq3-8}
\nonumber &u_i(x,t)=\\
\nonumber &=e^{-\lambda_i(t_{i+1}-t)}\left\{ \sum_{m=i}^{N-1}e^{-\sum_{k=i+1}^{m}\lambda_k(t_{k+1}-t_k)}\left[B^{+\quad\cdots~+~+}_{K_{i+1}\cdots K_m\frac{n}{R}}(x,t;t_{i+1},\cdots,t_m,t_{m+1})\right. \right. \\
\nonumber &\quad\quad\quad\quad\quad\quad\quad\quad\quad\quad\quad+\left.\frac{R}{n}A^{+\quad\cdots~+\quad-}_{K_{i+1}\cdots K_m\frac{n}{R}}(x,t;t_{i+1},\cdots,t_m,t_{m+1})\right]\\
\nonumber &\quad-\sum_{m=i}^{N-2}e^{-\sum_{k=i+1}^{m}\lambda_k(t_{k+1}-t_k)}B^{+\quad\cdots~+\quad+}_{K_{i+1}\cdots K_mK_{m+1}}(x,t;t_{i+1},\cdots,t_m,t_{m+1})\\
\nonumber&\quad+\sum_{m=i+1}^{N-1}\lambda_me^{-\sum_{k=i+1}^{m-1}\lambda_k(t_{k+1}-t_k)}\int_{t_m}^{t_{m+1}}e^{-\lambda_m(\tau-t_m)}\left[B^{+\quad\cdots~+~+}_{K_{i+1}\cdots K_m\frac{n}{R}}(x,t;t_{i+1},\cdots,t_m,\tau) \right.\\
\nonumber&\quad\quad\quad\quad\quad\quad\quad\quad\quad\quad+\left.\frac{R}{n}\left. A^{+\quad\cdots~+~~-}_{K_{i+1}\cdots K_m\frac{n}{R}}(x,t;t_{i+1},\cdots,t_m,\tau)\right]d\tau \right\}\\
&+\lambda_i\int_{t}^{t_{i+1}}e^{-\lambda_i(\tau-t)}\left[B^{+}_{\frac{n}{R}}(x,t;\tau; 0,b,s_V)+\frac{R}{n}A^{-}_{\frac{n}{R}}(x,t;\tau; 0,b,s_V)\right]d\tau.
\end{align}
{\it Here $B^{s_1\cdots s_m}_{K_1\cdots K_m}(x,t;t_1,\cdots,t_m)$ and $A^{s_1\cdots s_m}_{K_1\cdots K_m}(x,t;t_1,\cdots,t_m)$ are respectively the prices of m-th order bond and asset binaries with $0$-risk free rate, $b$-dividend rate and $s_V$-volatility}. (See lemma 1.)

    The proof is not difficult but somewhat complicated. We will prove it in the section 4. 

{\bf Remark 1}. In this theorem, the {\it financial meaning} of $u_i(x,t)$ is that it is the {\it relative price} of our bond in a subinterval {\it with respect to the risk free zero coupon bond}. We can derive the pricing formulae of our bond under other assumptions on the relations between  $K_i(i=1,\cdots,N)$ and $n/R$ using the same method.

\subsection{The Problem with Exogenous Recovery}

Instead of the assumption 5) let assume the following:

7) The default recovery $R_d$ is given as the form of exogenous face value 
\begin{equation}\label{eq3-9}
R_d =Re^{-r(T-t)}\quad (0\leq R\leq 1 \mathtt{~is~a~constant}.)
\end{equation}
Then under the assumptions 1), 2), 3), 4), 6) and 7) the pricing model of our bond is given as follows:
\begin{align}\label{eq3-10}
\nonumber &\frac{\partial C_i}{\partial t}+\frac{s_V^{2}}{2}V^2\frac{\partial^2 C_i}{\partial V^2}+(r-b)V\frac{\partial C_i}{\partial V}-(r+\lambda_i)C_i+\lambda_i Re^{-r(T-t)}=0,~t_i<t<t_{i+1}, \\
&C_i(t_{i+1})=C_{i+1}(t_{i+1})1(V>K_{i+1}e^{-r(T-t_{i+1})})+Re^{-r(T-t_{i+1})}1(V\leq K_{i+1}e^{-r(T-t_{i+1})}).
\end{align}
Here $i=0,1,\cdots,N-1$ and $C_N(V,  t)\equiv 1$.\\

{\bf Theorem 2}. (exogenous recovery) {\it Under the assumptions 1), 2), 3), 4), 6) and 7) the price of our bond, that is, the solution of \eqref{eq3-10} is represented as follows}: 
\begin{align}\label{eq3-11}
\nonumber C_i(V, t)=W_i(V/e^{-r(T-t)},&~t)e^{-r(T-t)}+[1-W_i(V/e^{-r(T-t)},~t)]Re^{-r(T-t)}, \\
& t_i\leq t<t_{i+1},~i=0,\cdots,N-1.
\end{align}
{\it Here}
\begin{align}\label{eq3-12}
W_i(x,t)=e^{-\lambda_i(t_{i+1}-t)-\sum_{k=i+1}^{N-1}\lambda_k(t_{k+1}-t_k)}B^{+\quad\cdots~+}_{K_{i+1}\cdots K_N}(x,t;t_{i+1},\cdots,t_N;0,b,s_V).
\end{align}
The proof is done by the same way with that of theorem 1. See the section 4.

{\bf Remark 2}. The {\it financial meaning} of the pricing formulae \eqref{eq3-11} is similar with that of \cite{OW}: the price of our defaultable bond at time $t$ can be seen as a probabilistic mean value of the current value $e^{-r(T-t)}$ of the bond in the case when there is no default after time $t$ and the value $Re^{-r(T-t)}$ of the bond in the case when default occurs after time $t$. So $W_i(V/e^{-r(T-t)},~t)$ is the survival probability after the time $t\in[t_i,t_{i+1})$, that is, the probability with which no default event occurs in the interval $(t,~T]$ and $1-W_i(V/e^{-r(T-t)},~t)$ is the ruin probability after the time $t\in[t_i,t_{i+1})$, that is, the probability with which default event occurs in the interval $(t,~T]$ when $t_i\leq t<t_{i+1}$. The formulae \eqref{eq3-11} can be written as follows:  
\begin{align}\label{eq3-13}
\nonumber C_i(V, t)=Re^{-r(T-t)}+&(1-R)W_i(V/e^{-r(T-t)},~t)e^{-r(T-t)}, \\
& t_i\leq t<t_{i+1},~i=0,\cdots,N-1.
\end{align}
The {\it financial meaning} of \eqref{eq3-13} is as follows: the {\bf first term} of \eqref{eq3-13} is the current price of the part to be given to bond holder {\it regardless of default occurs or not}, and the {\bf second term} is the {\it allowance} dependent on the survival probability after time $t$. If after some moment $t$, the default is certain ($W = 0$), then the price of the bond at $t$ is exactly the current price of default recovery $Re^{-r(T-t)}$. If the default recovery rate is zero, that is, $R = 0$, then the ratio of the defaultable bond price and default free zero coupon bond price is the very the survival probability after time $t$. If the default recovery rate is full, that is, $R = 1$, then default event does not effect to the bond price and defaultable bond price is the same with default free zero coupon bond price. 

\subsection{Illustration of the Effect of Parameters on the Bond Price}
In this subsection we illustrate the effect of several parameters including recovery rate $R$, volatility $s_V$ of firm value, $x = V/e^{-r(T-t)}$ (relative price of firm value), default boundary $K$ and default intensity $\lambda$ on the price of the defaultable bonds.  Let $N=2,~t_1=3, t_2=6$ (annum).

Basic data for calculation are as follows: Short rate $r = 0.1$; Firm value: dividend rate $b = 0.05$, volatility $s_V =1.0$, $x = V/e^{-r(T-t)}= 200$; $\lambda_0=0.002,~\lambda_1=0.005$ are respectively default intensities in the intervals $(0,t_1)$, $(t_1,t_2)$; $K_1=K_2=100$ is default barrier at time $t_1,~t_2$; recovery rate $R = 0.5$.

We will analyze $(t, C)$-plot changing one of $R, s_V, x, K$ and $\lambda$ under keeping the remainder of data on as the above. See the following figures \ref{fig01}--\ref{fig09}.
\begin{figure}[h]
   \centering
   \includegraphics[width=0.63\textwidth]{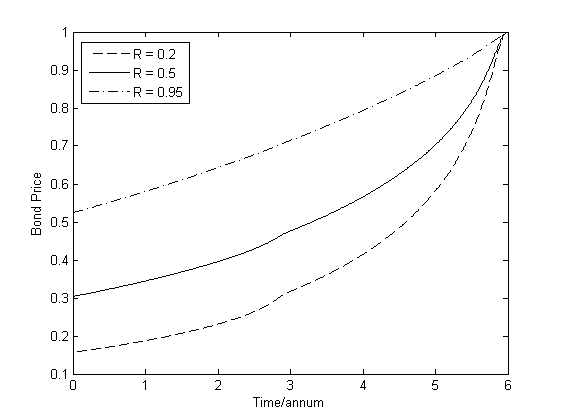}
   \caption{Plot $(t : C)$ when $R= 0.2, 0.5, 0.95$}
  \label{fig01}
\end{figure}
\begin{figure}[h]
   \centering
   \includegraphics[width=0.87\textwidth]{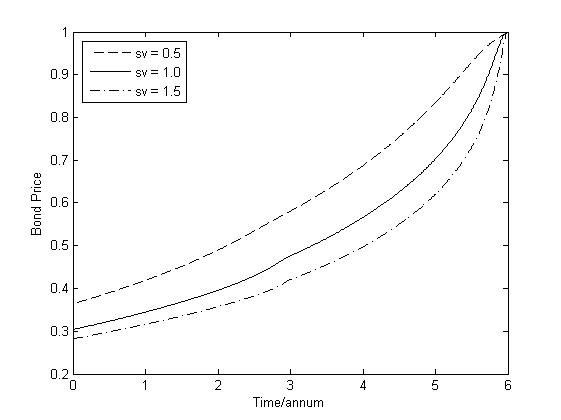}
   \caption{Plot $(t : C)$ when $s_V = 0.5, 1.0, 1.5$}
  \label{fig02}
\end{figure}
\begin{figure}[h]
   \centering
   \includegraphics[width=0.87\textwidth]{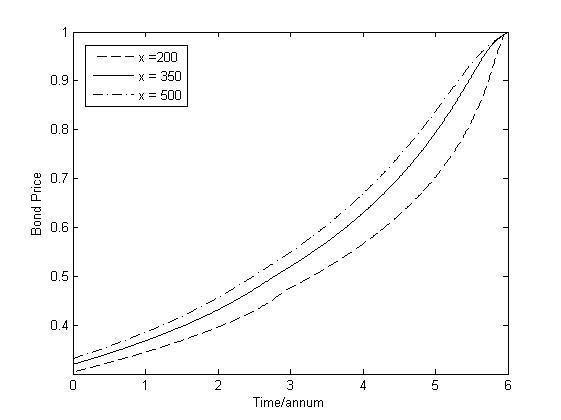}
   \caption{Plot $(t : C)$ when $x = V/e^{-r(T-t)}= 200, 350, 500$}
  \label{fig03}
\end{figure}
\begin{figure}[h]
   \centering
   \includegraphics[width=0.9\textwidth]{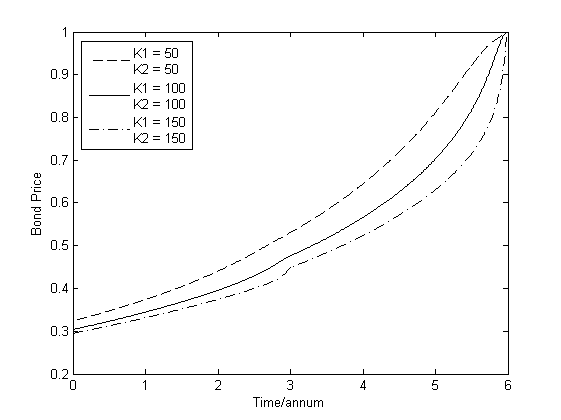}
   \caption{Plot $(t : C)$ when $(K_1,K_2)=(50,50), (100,100), (150, 150)$}
  \label{fig04}
\end{figure}
\begin{figure}[h]
   \centering
   \includegraphics[width=0.89\textwidth]{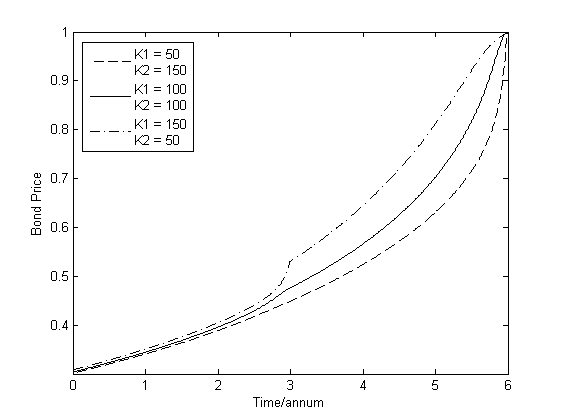}
   \caption{Plot $(t : C)$ when $(K_1,K_2)=(50,150), (100,100), (150, 50)$}
  \label{fig05}
\end{figure}
\begin{figure}[h]
   \centering
   \includegraphics[width=0.9\textwidth]{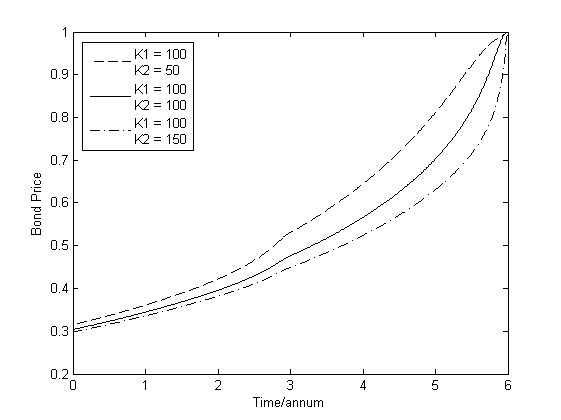}
   \caption{Plot $(t : C)$ when $K_1=100, K_2 =50,100, 150$}
  \label{fig06}
\end{figure}
\begin{figure}[h]
   \centering
   \includegraphics[width=0.89\textwidth]{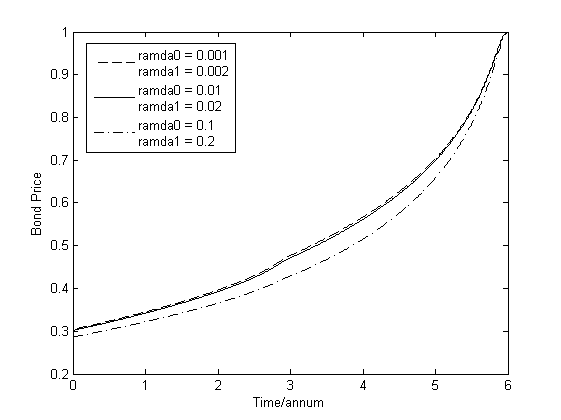}
   \caption{Plot $(t : C)$ when $(\lambda_0,\lambda_1)=(0.001,0.002), (0.01,0.02), (0.1, 0.2)$}
  \label{fig07}
\end{figure}
\begin{figure}[h]
   \centering
   \includegraphics[width=0.9\textwidth]{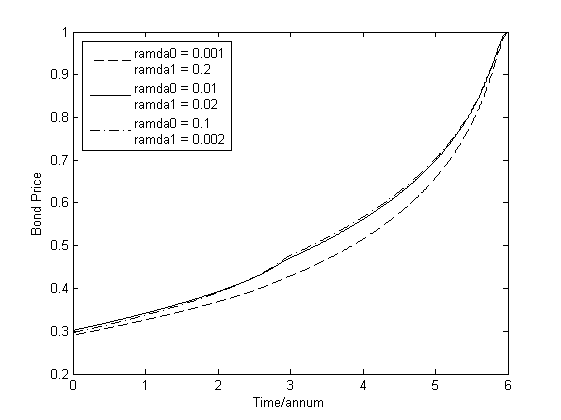}
   \caption{Plot $(t : C)$ when $(\lambda_0,\lambda_1)=(0.001,0.2), (0.01,0.02), (0.1, 0.002)$}
  \label{fig08}
\end{figure}
\begin{figure}[h]
   \centering
   \includegraphics[width=0.89\textwidth]{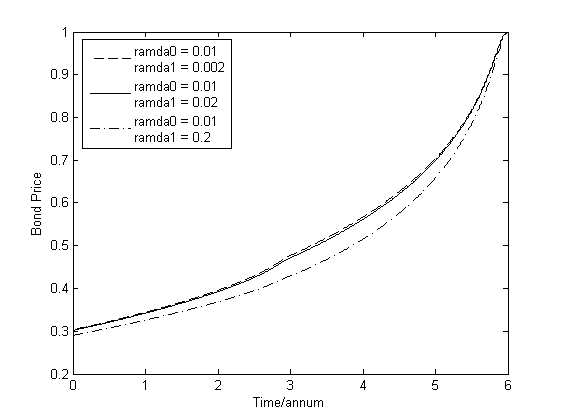}
   \caption{Plot $(t : C)$ when $\lambda_0=0.01, \lambda_1=0.002,0.02,0.2$}
  \label{fig09}
\end{figure}

Note that figure \ref{fig01} shows that increase of recovery rate results in increase of bond price. Figure \ref{fig02} shows that increase of volatility of firm value results in decrease of bond price. The reason is that when $s_V$ increases, the firm value fluctuates more seriously and there are more risks of default, which results in decrease of bond price. Figure \ref{fig03} shows that increase of firm value results in increase of bond price. Figures \ref{fig04}--\ref{fig09} show the effect of default barrier and default intensity on bond price. In particular, in the figure \ref{fig05} (or \ref{fig08}) we can see the mixed effect of increase of $K_1$ (or $\lambda_0$) and decrease of $K_2$ (or $\lambda_1$) in the subinterval $[0, 3]$.

\subsection{Credit Spread Analysis}

In this subsection, we illustrate the effect of several parameters including recovery rate $R$, volatility $s_V$ of firm value, $x=V/e^{-r(T-t)}$, default boundary $K$ and default intensity $\lambda$ on credit spreads. The {\it credit spread} is defined using the difference between the yields of the defaultable bond $C$ and the default-free bond $e^{-r(T-t)}$ and is given by the following expression:
\begin{equation*}
CS =-\frac{\ln[C/e^{-r(T-t)}]}{T-t}.
\end{equation*}
 
For simplicity, we only consider the case of exogenous default recovery (theorem 2). Then, the credit spread is differently given in every subinterval as follows: 
\begin{equation} \label{eq3-14}
CS_i=-\frac{\ln[R+(1-R)W_i(V/e^{-r(T-t)},t)]}{T-t}, t_i\leq t<t_{i+1},i=0,\cdots,N-1.
\end{equation}
Let $N=2, t_1=3, t_2=T=6$ (annum) as in the above.

Basic data for calculation of $CS$ are as follows: Short rate $r = 0.1$; Firm value: dividend rate $b = 0.05$, volatility $s_V =1.0$, $x = V/e^{-r(T-t)}=200$; $\lambda_0=0.002,~\lambda_1=0.005$ are respectively default intensities in the intervals $(0,t_1)$, $(t_1,t_2)$; $K_1=K_2=100$ is default barrier at time $t_1,~t_2$; recovery rate $R = 0.5$.

\begin{figure}[h]
   \centering
   \includegraphics[width=0.9\textwidth]{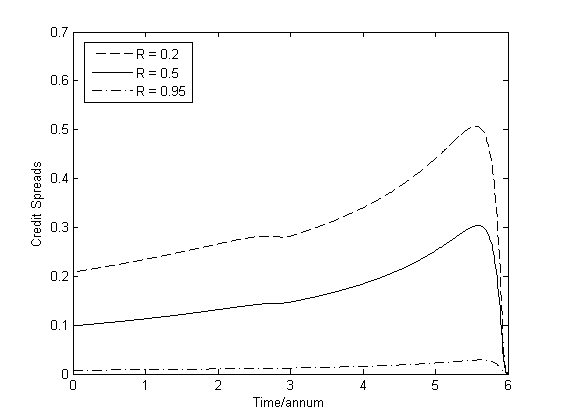}
   \caption{Plot $(t : CS)$ when $R= 0.2, 0.5, 0.95$}
  \label{fig10}
\end{figure}

\begin{figure}[h]
   \centering
   \includegraphics[width=0.89\textwidth]{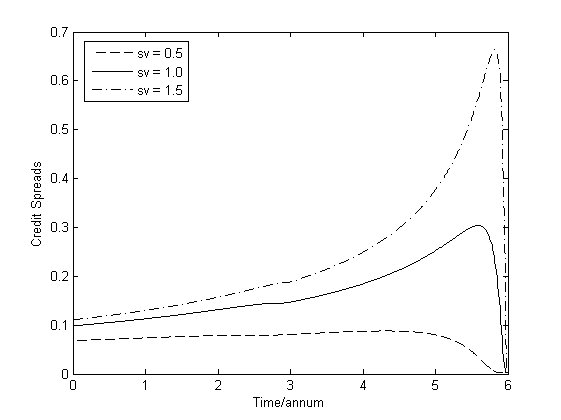}
   \caption{Plot $(t : CS)$ when $s_V = 0.5, 1.0, 1.5$}
  \label{fig11}
\end{figure}

\begin{figure}[h]
   \centering
   \includegraphics[width=0.89\textwidth]{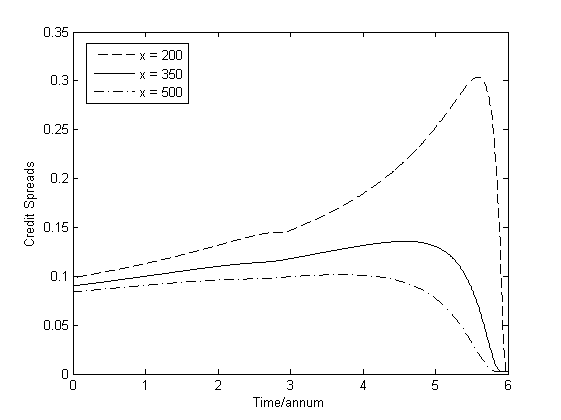}
   \caption{Plot $(t : CS)$ when $x = V/e^{-r(T-t)}= 200, 350, 500$}
  \label{fig12}
\end{figure}

\begin{figure}[h]
   \centering
   \includegraphics[width=0.9\textwidth]{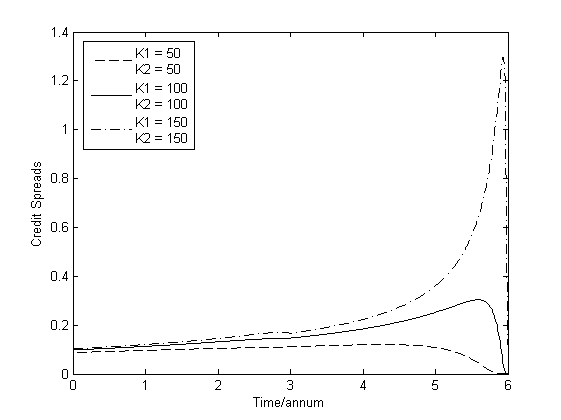}
   \caption{Plot $(t : CS)$ when $(K_1,K_2)=(50,50), (100,100), (150, 150)$}
  \label{fig13}
\end{figure}

\begin{figure}[h]
   \centering
   \includegraphics[width=0.89\textwidth]{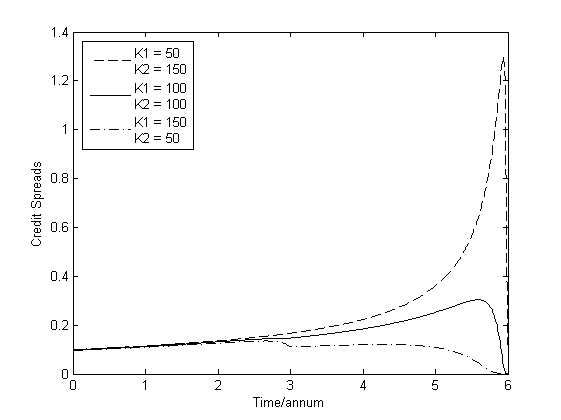}
   \caption{Plot $(t : CS)$ when $(K_1,K_2)=(50,150), (100,100), (150, 50)$}
  \label{fig14}
\end{figure}

\begin{figure}[h]
   \centering
   \includegraphics[width=0.9\textwidth]{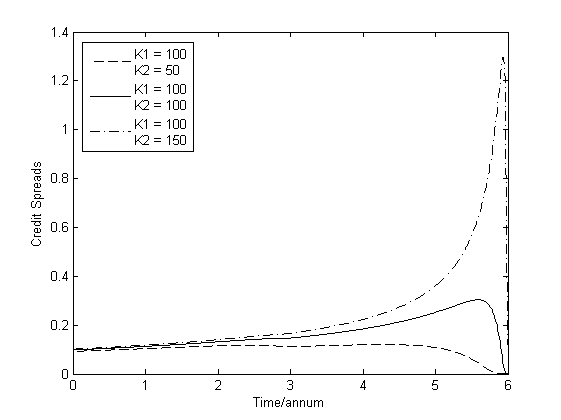}
   \caption{Plot $(t : CS)$ when $K_1=100, K_2 =50,100, 150$}
  \label{fig15}
\end{figure}

\begin{figure}[h]
   \centering
   \includegraphics[width=0.89\textwidth]{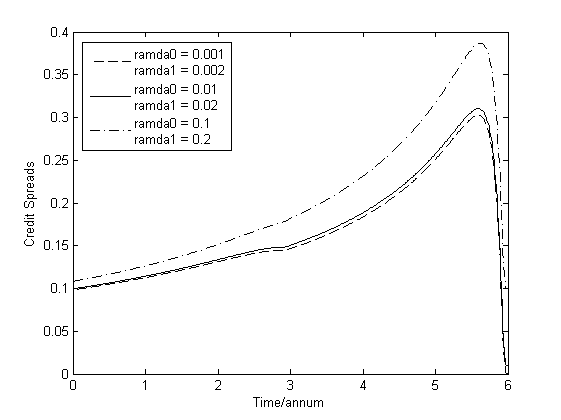}
   \caption{\small Plot$(t:CS)$ when $(\lambda_0,\lambda_1)=(0.001,0.002),(0.01,0.02),(0.1,0.2)$}
  \label{fig16}
\end{figure}

\begin{figure}[h]
   \centering
   \includegraphics[width=0.9\textwidth]{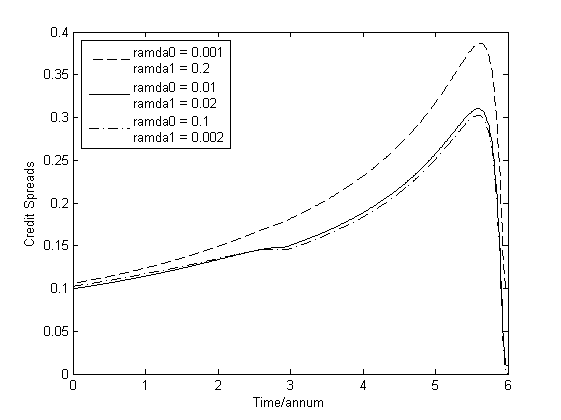}
   \caption{\small Plot$(t:CS)$ when $(\lambda_0,\lambda_1)=(0.001,0.2),(0.01,0.02),(0.1, 0.002)$}
  \label{fig17}
\end{figure}

\begin{figure}[h]
   \centering
   \includegraphics[width=0.89\textwidth]{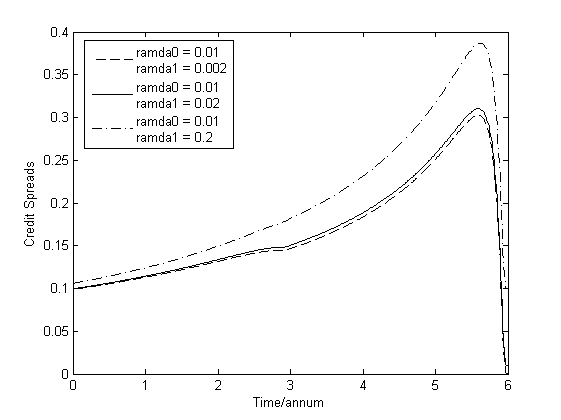}
   \caption{\small Plot $(t : CS)$ when $\lambda_0=0.01, \lambda_1=0.002,0.02,0.2$}
  \label{fig18}
\end{figure}

We will analyze $(t , CS)$-plot changing one of $R, s_V, x, K$ and $\lambda$ under keeping the remainder of data on as the above. In what follows, the figure \ref{fig10} shows that increase of recovery rate results in decrease of credit spread. Figure \ref{fig11} shows that increase of volatility of firm value results in increase of credit spread. The reason is that when $s_V$ increases, the firm value fluctuates more seriously and there are more risks of default, which results in increase of credit spread. Figure \ref{fig12} shows that increase of firm value results in decrease of credit spread. Figures \ref{fig13}--\ref{fig18} show the effect of default barrier and default intensity on credit spread. In particular, in the figure \ref{fig14} (or \ref{fig16}) we can see the mixed effect of increase of $K_1$ (or $\lambda_0$) and decrease of $K_2$ (or $\lambda_1$) in the subinterval $[0, 3]$.


\section{The Proofs of The Pricing Formulae}

{\bf The Proof of Theorem 1}. i) In \eqref{eq3-5}, we use change of numeraire
\begin{equation} \label{eq4-1}
x=\frac{V}{e^{-r(T-t)}},~u_i(x,t)=-\frac{C_i(V,t)}{e^{-r(T-t)}}, t_i\leq t<t_{i+1},i=0,\cdots,N-1.
\end{equation}
Substituting \eqref{eq4-1} into \eqref{eq3-5} we get
\begin{align}\label{eq4-2}
\nonumber &\frac{\partial u_i}{\partial t}+\frac{s_V^{2}}{2}x^2\frac{\partial^2 u_i}{\partial x^2}-bx\frac{\partial u_i}{\partial x}-\lambda_i u_i+\lambda_i\min\{1, \frac{R}{n}x\}=0,~t_i<t<t_{i+1}, x>0,\\
&u_i(x,t_{i+1})=u_{i+1}(x,t_{i+1})1(x>K_{i+1})+\min\{1, \frac{R}{n}x\}1(x\leq K_{i+1}),i=0,\cdots,N-1.
\end{align}
Here $u_N(x,t)\equiv 1$. From the assumption
\begin{equation} \label{eq4-3}
K_i\leq n/R, i=1,\cdots,N
\end{equation}
If $V(t_i)\leq K_ie^{-r(T-t_i)}$, that is, if the default event occurs at time $t_i$, then $\min\{e^{-r(T-t_i)}, RV(t_i)/n\}=RV(t_i)/n$ and we have
\begin{equation} \label{eq4-4}
\min\{1, \frac{R}{n}x\}1(x\leq K_{i+1})=\frac{R}{n}x\cdot 1(x\leq K_{i+1}).
\end{equation}
Then the problem \eqref{eq4-2} is changed into the following one.
\begin{align}\label{eq4-5}
\nonumber &\frac{\partial u_i}{\partial t}+\frac{s_V^{2}}{2}x^2\frac{\partial^2 u_i}{\partial x^2}-bx\frac{\partial u_i}{\partial x}-\lambda_i u_i+\lambda_i\min\{1, \frac{R}{n}x\}=0,~t_i<t<t_{i+1}, x>0,\\
&u_i(x,t_{i+1})=u_{i+1}(t_{i+1})1(x>K_{i+1})+\frac{R}{n}x\cdot 1(x\leq K_{i+1}),i=0,\cdots,N-1.
\end{align}
When $i=N-1$, \eqref{eq4-5} is as follows: 
\begin{align}\label{eq4-6}
\nonumber &\frac{\partial u_{N-1}}{\partial t}+\frac{s_V^{2}}{2}x^2\frac{\partial^2 u_{N-1}}{\partial x^2}-bx\frac{\partial u_{N-1}}{\partial x}-\lambda_{N-1} u_{N-1}+\lambda_{N-1}\min\{1, \frac{R}{n}x\}=0,\\
\nonumber &\qquad\qquad\qquad\qquad\qquad\qquad\qquad\qquad\qquad\qquad\qquad\qquad~t_{N-1}<t<T, x>0,\\
&u_{N-1}(x,T)=1(x>K_N)+\frac{R}{n}x\cdot 1(x\leq K_N),~~\qquad x>0.
\end{align}
This is a terminal value problem for an inhomogenous Black-Scholes equation with coefficients $r=\lambda_{N-1},~q=\lambda_{N-1}+b,~\sigma=s_V$. Let $L_{N-1}$ be the Black-Scholes partial differential operator with coefficients $r=\lambda_{N-1},~q=\lambda_{N-1}+b, \sigma=s_V$, that is, 
\[L_{N-1}u=\frac{\partial u}{\partial t}+\frac{s_V^{2}}{2}x^2\frac{\partial^2 u}{\partial x^2}-bx\frac{\partial u}{\partial x}-\lambda_{N-1}u.\]
Then the solution of \eqref{eq4-6} is provided by sum of the solutions $U_1$ and $U_2$ to the two following problems:
\begin{align}\label{eq4-7}
\nonumber &L_{N-1}U_1=0,\qquad\qquad\qquad\qquad\quad~t_{N-1}<t<T, x>0,\\
&U_1(x,T)=1(x>K_N)+\frac{R}{n}x\cdot 1(x\leq K_N),\qquad x>0.
\end{align}
\begin{align}\label{eq4-8}
\nonumber &L_{N-1}U_2+\lambda_{N-1}\min\{1, \frac{R}{n}x\}=0,\quad t_{N-1}<t<T, x>0,\\
&U_2(x,T)=0,\qquad\qquad\qquad\qquad\qquad\qquad\qquad x>0.
\end{align}
The terminal payoff of \eqref{eq4-7} is linear combination of the terminal payoffs of bond and asset binaries (refer to section 2) and thus the solution to \eqref{eq4-7} is given as follows:
\[U_1=B_{K_N}^+(x,t;T;\lambda_{N-1},\lambda_{N-1}+b,s_V)+\frac{R}{n}A_{K_N}^-(x,t;T;\lambda_{N-1},\lambda_{N-1}+b,s_V),~t_{N-1}\leq t<T.\]
The problem \eqref{eq4-8} is a 0-terminal value problem of an inhomogeneous equation and thus we can use the {\it Duhamel's principle} to solve it. Fix $\tau\in(t_{N-1},T]$ and let $W(x,t;\tau)$ be the solution to the following terminal value problem:
\begin{align*}
&L_{N-1}W=0,\qquad\qquad\qquad t_{N-1}<t<\tau, x>0,\\
&W(x,\tau;\tau)=\lambda_{N-1}\min\{1, \frac{R}{n}x\},\qquad x>0.
\end{align*}
Since $\lambda_{N-1}\min\{1, \frac{R}{n}x\}=\lambda_{N-1}\left[1(x>n/R)+\frac{R}{n}x\cdot 1(x<n/R)\right]$, the solution is given as follows:
\[W(x,t;\tau)=\lambda_{N-1}\left[B_{n/R}^+(x,t;\tau;\lambda_{N-1},\lambda_{N-1}+b,s_V)\right.+\]
\[\qquad\qquad\qquad+\frac{R}{n}\left.A_{n/R}^-(x,t;\tau;\lambda_{N-1},\lambda_{N-1}+b,s_V)\right],~t_{N-1}\leq t<\tau,~x>0.\]
Then the solution $U_2$ to \eqref{eq4-8} is given as follows:
\begin{align*}
U_2&=\int_{t}^{T}W(x,t;\tau)d\tau=\\
&=\lambda_{N-1}\int_{t}^{T}\left[B_{n/R}^+(x,t;\tau;\lambda_{N-1},\lambda_{N-1}+b,s_V)\right.+\\
&\qquad+\frac{R}{n}\left.A_{n/R}^-(x,t;\tau;\lambda_{N-1},\lambda_{N-1}+b,s_V)\right]d\tau,~t_{N-1}\leq t<T,~x>0.
\end{align*}
Thus the solution to \eqref{eq4-6} is provided by $u_{N-1}(x,t)=U_1+U_2$, that is, 
\begin{align}\label{eq4-9}
\nonumber u&_{N-1}(x,t)=\\
\nonumber &=B_{K_N}^+(x,t;T;\lambda_{N-1},\lambda_{N-1}+b,s_V)+\frac{R}{n}A_{K_N}^-(x,t;T;\lambda_{N-1},\lambda_{N-1}+b,s_V)+\\
\nonumber &+\lambda_{N-1}\int_{t}^{T}\left[B_{n/R}^+(x,t;\tau;\lambda_{N-1},\lambda_{N-1}+b,s_V)\right.+\\
&\quad\qquad+\frac{R}{n}\left.A_{n/R}^-(x,t;\tau;\lambda_{N-1},\lambda_{N-1}+b,s_V)\right]d\tau,~t_{N-1}\leq t<T,~x>0.
\end{align}
For our further purpose, using the relations \eqref{eq2-11} we rewrite \eqref{eq4-9} by the price of bond and asset binaries with the coefficients $r=0, q=b, \sigma=s_V$: 
\begin{align}\label{eq4-10}
\nonumber u&_{N-1}(x,t)=e^{-\lambda_{N-1}(T-t)}\left[B_{K_N}^+(x,t;T;0,b,s_V)+\frac{R}{n}A_{K_N}^-(x,t;T;0,b,s_V)\right]+\\
\nonumber &+\lambda_{N-1}\int_{t}^{T}e^{-\lambda_{N-1}(\tau-t)}\left[B_{n/R}^+(x,t;\tau;0,b,s_V)+\frac{R}{n}A_{n/R}^-(x,t;\tau;0,b,s_V)\right]d\tau,\\
&\qquad\qquad\qquad\qquad\qquad\qquad\qquad\qquad~t_{N-1}\leq t<T,~x>0.
\end{align}
Now solve \eqref{eq4-5} when $i=N-2$. In this case \eqref{eq4-5} is as follows:
\begin{align}\label{eq4-11}
\nonumber &\frac{\partial u_{N-2}}{\partial t}+\frac{s_V^{2}}{2}x^2\frac{\partial^2 u_{N-2}}{\partial x^2}-bx\frac{\partial u_{N-2}}{\partial x}-\lambda_{N-2} u_{N-1}+\lambda_{N-2}\min\{1, \frac{R}{n}x\}=0,\\
\nonumber &\quad\qquad\qquad\qquad\qquad\qquad\qquad\qquad\qquad\qquad\qquad\qquad~t_{N-2}<t<t_{N-1}, x>0,\\
&u_{N-2}(x,t_{N-1})=u_{N-1}(x,t_{N-1})1(x>K_{N-1})+\frac{R}{n}x\cdot 1(x\leq K_{N-1}).
\end{align}
This is a terminal value problem of the inhomogeneous Black-Scholes equation with coefficients $r=\lambda_{N-2},~q=\lambda_{N-2}+b,~\sigma=s_V$.

{\bf Remark 3}. If we consider \eqref{eq4-9}, then the expiry payoff of \eqref{eq4-11} is the linear combination of first order binaries or zero and integrals of first order binaries or zero and therefore you could think that it is natural to solve \eqref{eq4-11} using the pricing formulae of second order binaries and their integrals. But we must {\it note} that the coefficients of \eqref{eq4-11} {\it are different} from those of \eqref{eq4-6} and \eqref{eq4-9} and thus we can't directly apply the pricing formulae of second order binaries here. Fortunately, the {\it differences} between risk free rates and dividend rates in adjacent subintervals are {\it all a constant} $-b$ and {\it volatility is not changed} in whole time interval and thus we can carefully use the pricing formulae of second order binaries with \eqref{eq2-11} together to give a representation of the solution to \eqref{eq4-11}.  

If we rewrite the terminal payoff of \eqref{eq4-11} into prices of binaries with the coefficients $r=\lambda_{N-2},~q=\lambda_{N-2}+b,~\sigma=s_V$ using \eqref{eq2-11}, then from \eqref{eq4-9} we get:
\begin{align*}
u_{N-1}&(x,t_{N-1})=e^{-(\lambda_{N-1}-\lambda_{N-2})(T-t_{N-1})}\left[B_{K_N}^+(x,t;T;\lambda_{N-2},\lambda_{N-2}+b,s_V)\right.+\\
&\quad\qquad\qquad\qquad\qquad\qquad\qquad+\frac{R}{n}\left.A_{K_N}^-(x,t;T;\lambda_{N-2},\lambda_{N-2}+b,s_V)\right]+\\
&+\lambda_{N-1}\int_{t}^{T}e^{-(\lambda_{N-1}-\lambda_{N-2})(\tau-t_{N-1})}\left[B_{n/R}^+(x,t;\tau;\lambda_{N-2},\lambda_{N-2}+b,s_V)\right.+\\
&\qquad\qquad\qquad\qquad\qquad\qquad+\frac{R}{n}\left.A_{n/R}^-(x,t;\tau;\lambda_{N-2},\lambda_{N-2}+b,s_V)\right]d\tau.
\end{align*}
Let $L_{N-2}$ be the Black-Scholes partial differential operator with coefficients $r=\lambda_{N-2},~q=\lambda_{N-2}+b,~\sigma=s_V$. Then the solution to \eqref{eq4-11} is the sum $U_1+U_2+U_3$ of the solutions to the following three problems:
\begin{align}\label{eq4-12}
\nonumber L_{N-2}U_1=0&,\qquad\qquad\qquad\qquad\qquad\qquad\quad~t_{N-2}<t<t_{N-1}, x>0,\\
\nonumber U_1(x,t_{N-1})&=e^{-(\lambda_{N-1}-\lambda_{N-2})(T-t_{N-1})}\left[B_{K_N}^+(x,t_{N-1};T;\lambda_{N-2},\lambda_{N-2}+b,s_V)\right.+\\
\nonumber &\qquad\quad+\frac{R}{n}\left.A_{K_N}^-(x,t_{N-1};T;\lambda_{N-2},\lambda_{N-2}+b,s_V)\right]1(x>K_{N-1})+\\
&+\frac{R}{n}x\cdot 1(x\leq K_{N-1}),
\end{align}
\begin{align}\label{eq4-13}
\nonumber &L_{N-2}U_2=0,\qquad\qquad\qquad\qquad\qquad\qquad\quad~t_{N-2}<t<t_{N-1}, x>0,\\
\nonumber &U_2(x,t_{N-1})=\\
\nonumber &=\lambda_{N-1}\int_{t_{N-1}}^{T}e^{-(\lambda_{N-1}-\lambda_{N-2})(\tau-t_{N-1})}\left[B_{\frac{n}{R}}^+(x,t_{N-1};\tau;\lambda_{N-2},\lambda_{N-2}+b,s_V)\right.+\\
&\qquad\quad+\frac{R}{n}\left.A_{\frac{n}{R}}^-(x,t_{N-1};\tau;\lambda_{N-2},\lambda_{N-2}+b,s_V)\right]d\tau\cdot 1(x>K_{N-1}).
\end{align}
\begin{align}\label{eq4-14}
\nonumber &L_{N-2}U_3+\lambda_{N-2}\min\{1, \frac{R}{n}x\}=0,\qquad\qquad\qquad t_{N-2}<t<t_{N-1}, x>0,\\
&U_3(x,t_{N-1})=0,\qquad\qquad\qquad\qquad\qquad\qquad\qquad\qquad\qquad\quad x>0.
\end{align}
Using the prices of first and second order binaries \eqref{eq2-6} and \eqref{eq2-7}, the solution to \eqref{eq4-12} is given as follows:
\begin{align}\label{eq4-15}
\nonumber U_1&(x,t)=\\
\nonumber &=e^{-(\lambda_{N-1}-\lambda_{N-2})(T-t_{N-1})}\left[B_{K_{N-1}K_N}^{+\quad+}(x,t;t_{N-1},T;\lambda_{N-2},\lambda_{N-2}+b,s_V)\right.+\\
\nonumber &\qquad\qquad\qquad\qquad\quad+\frac{R}{n}\left.A_{K_{N-1}K_N}^{+\quad -}(x,t;t_{N-1},T;\lambda_{N-2},\lambda_{N-2}+b,s_V)\right]+\\
&+\frac{R}{n}A_{K_{N-1}}^{-}(x,t;t_{N-1};\lambda_{N-2},\lambda_{N-2}+b,s_V), ~t_{N-2}\leq t<t_{N-1}.
\end{align}
From the corollary of Lemma 2, the solution to \eqref{eq4-13} is given as follows:
\begin{align}\label{eq4-16}
\nonumber &U_2(x,t)=\\
\nonumber &=\lambda_{N-1}\int_{t_{N-1}}^Te^{-(\lambda_{N-1}-\lambda_{N-2})(\tau-t_{N-1})}\left[B_{K_{N-1}\frac{n}{R}}^{~+\quad +}(x,t,t_{N-1},\tau;\lambda_{N-2},\lambda_{N-2}+b,s_V)\right.+\\
&\qquad+\frac{R}{n}\left.A_{K_{N-1}\frac{n}{R}}^{~+\quad -}(x,t;t_{N-1},\tau;\lambda_{N-2},\lambda_{N-2}+b,s_V)\right]d\tau,~t_{N-2}\leq t<t_{N-1}.
\end{align}
\eqref{eq4-14} is a 0-terminal value problem of an inhomogeneous equation just like \eqref{eq4-8}, so its solution is given by
\begin{align}\label{eq4-17}
\nonumber &U_3(x,t)=\lambda_{N-2}\int_{t}^{t_{N-1}}\left[B_{n/R}^+(x,t;\tau;\lambda_{N-2},\lambda_{N-2}+b,s_V)\right.+\\
&\qquad+\frac{R}{n}\left.A_{n/R}^-(x,t;\tau;\lambda_{N-2},\lambda_{N-2}+b,s_V)\right]d\tau,~t_{N-2}\leq t<t_{N-1}.
\end{align}
Thus we obtain the representation of $u_{N-2}(x,t)=U_1+U_2+U_3$, that is,     
\begin{align}\label{eq4-18}
\nonumber &u_{N-2}(x,t)=\\
\nonumber &=e^{-(\lambda_{N-1}-\lambda_{N-2})(T-t_{N-1})}\left[B_{K_{N-1}K_N}^{+\quad+}(x,t;t_{N-1},T;\lambda_{N-2},\lambda_{N-2}+b,s_V)\right.+\\
\nonumber &\qquad\qquad\qquad\qquad\quad+\frac{R}{n}\left.A_{K_{N-1}K_N}^{+\quad -}(x,t;t_{N-1},T;\lambda_{N-2},\lambda_{N-2}+b,s_V)\right]+\\
\nonumber &+\frac{R}{n}A_{K_{N-1}}^{-}(x,t;t_{N-1};\lambda_{N-2},\lambda_{N-2}+b,s_V) \\
\nonumber &+\lambda_{N-1}\int_{t_{N-1}}^Te^{-(\lambda_{N-1}-\lambda_{N-2})(\tau-t_{N-1})}\left[B_{K_{N-1}\frac{n}{R}}^{~+\quad +}(x,t,t_{N-1},\tau;\lambda_{N-2},\lambda_{N-2}+b,s_V)\right.+\\
\nonumber &\qquad\qquad\qquad\qquad\quad+\frac{R}{n}\left.A_{K_{N-1}\frac{n}{R}}^{~+\quad -}(x,t;t_{N-1},\tau;\lambda_{N-2},\lambda_{N-2}+b,s_V)\right]d\tau \\
\nonumber &+\lambda_{N-2}\int_{t}^{t_{N-1}}\left[B_{n/R}^+(x,t;\tau;\lambda_{N-2},\lambda_{N-2}+b,s_V)\right.+\\
&\qquad\quad+\frac{R}{n}\left.A_{n/R}^-(x,t;\tau;\lambda_{N-2},\lambda_{N-2}+b,s_V)\right]d\tau,~t_{N-2}\leq t<t_{N-1}.
\end{align}
For our further purpose, using the relations \eqref{eq2-11} we rewrite \eqref{eq4-18} by the price of bond and asset binaries with the coefficients $r=0,~q=b,~\sigma=s_V$ to get
\begin{align}\label{eq4-19}
\nonumber &u_{N-2}(x,t)=\\
\nonumber &=e^{-\lambda_{N-2}(T-t)-(\lambda_{N-1}-\lambda_{N-2})(T-t_{N-1})}\left[B_{K_{N-1}K_N}^{+\quad+}(x,t;t_{N-1},T;0,b,s_V)\right.+\\
\nonumber &\qquad\qquad\qquad\qquad\qquad\qquad\qquad+\frac{R}{n}\left.A_{K_{N-1}K_N}^{+\quad -}(x,t;t_{N-1},T;0,b,s_V)\right]+\\
\nonumber &+e^{-\lambda_{N-2}(t_{N-1}-t)}\frac{R}{n}A_{K_{N-1}}^{-}(x,t;t_{N-1};0,b,s_V)+ \\
\nonumber &+\lambda_{N-1}\int_{t_{N-1}}^Te^{-\lambda_{N-2}(\tau-t)-(\lambda_{N-1}-\lambda_{N-2})(\tau-t_{N-1})}\left[B_{K_{N-1}\frac{n}{R}}^{~+\quad +}(x,t,t_{N-1},\tau;0,b,s_V)\right.+\\
\nonumber &\qquad\qquad\qquad\qquad\qquad\qquad\qquad+\frac{R}{n}\left.A_{K_{N-1}\frac{n}{R}}^{~+\quad -}(x,t;t_{N-1},\tau;0,b,s_V)\right]d\tau+\\
\nonumber &+\lambda_{N-2}\int_{t}^{t_{N-1}}e^{-\lambda_{N-2}(\tau-t)}\left[B_{\frac{n}{R}}^+(x,t;\tau;0,b,s_V)\right.+\frac{R}{n}\left.A_{\frac{n}{R}}^-(x,t;\tau;0,b,s_V)\right]d\tau,\\
&\qquad\qquad\qquad\qquad\qquad\qquad\qquad\qquad\qquad\qquad t_{N-2}\leq t<t_{N-1}.
\end{align}

By induction we can obtain the representations of all $u_i(x,t)(i=0,\cdots,N-1)$. If in every representation of $u_i(x,t)$ we replace the higher order binaries with the coefficients $r=\lambda_i,~q=\lambda_i+b,~\sigma=s_V$ into the higher order binaries with the coefficients $r=0,~q=b,~\sigma=s_V$ using the relation \eqref{eq2-11} and arrange the exponents properly, we soon obtain \eqref{eq3-7}. If we return to the original variable $V$ and the unknown function $C$ using \eqref{eq4-1}, then we soon obtain \eqref{eq3-6}. 

   The proof of ii) is the same. (QED)\\

{\bf The proof of Theorem 2}. In \eqref{eq3-10}, if we use change of numeraire \eqref{eq4-1}, then we have
\begin{align}\label{eq4-20}
\nonumber &\frac{\partial u_i}{\partial t}+\frac{s_V^{2}}{2}x^2\frac{\partial^2 u_i}{\partial x^2}-bx\frac{\partial u_i}{\partial x}-\lambda_i u_i+\lambda_i R=0,~t_i<t<t_{i+1}, x>0,\\
&u_i(x,t_{i+1})=u_{i+1}(x,t_{i+1})1(x>K_{i+1})+R\cdot 1(x\leq K_{i+1}),i=0,\cdots,N-1.
\end{align}
Here $u_N(x,t)\equiv 1$. We use the change of unknown function 
\begin{equation}\label{eq4-21}
u_i=(1-R)W_i+R, i=0,\cdots,N-1.
\end{equation}
Then the problem \eqref{eq4-20} is changed into the following one.
\begin{align}\label{eq4-22}
\nonumber &\frac{\partial W_i}{\partial t}+\frac{s_V^{2}}{2}x^2\frac{\partial^2 W_i}{\partial x^2}-bx\frac{\partial W_i}{\partial x}-\lambda_i W_i=0,~t_i<t<t_{i+1}, x>0,\\
&W_i(x,t_{i+1})=W_{i+1}(x,t_{i+1})1(x>K_{i+1}),~x>0,~i=0,\cdots,N-1.
\end{align}
Here  $W_N(x,t)\equiv 1$. These equations are simpler than ones in theorem 1 (note that \eqref{eq4-22} are {\it homogenous} Black-Scholes equations) and we can easily solve them with the same method in the above to get \eqref{eq3-12} and \eqref{eq3-11}.

\section{Conclusions}

In this paper we studied the pricing of defaultable bond with discrete default intensity and barrier under constant risk free short rate using higher order binary options (\cite{buc, OK1, OK2}) and their integrals. We considered both {\it endogenous} and {\it exogenous} default recovery. Our pricing problem is derived to a solving problem of {\it inhomogeneous} or {\it homogeneous Black-Scholes PDEs} with {\it different coefficients} and terminal value of binary type in every subinterval between the two adjacent announcing dates. See \eqref{eq3-5} and \eqref{eq3-10}. In order to deal with the difference of coefficients in subintervals we used a {\it relation} \eqref{eq2-11} between prices of higher order binaries with different coefficients. In our model, due to the inhomogenous term related to endogenous recovery, our bond prices are represented by {\it not only} the prices of higher binary options {\it but also} the {\it integrals} of them. See the formulae \eqref{eq3-7} and \eqref{eq3-8}(3.8). So first we provided the pricing formulae (corollary of lemma 2) of a {\it special binary option} called {\it integral of i-th binary or nothing} and then we obtain the pricing formulae of our defaultable corporate bond by using the pricing formulae of higher binary options and integrals of them and provided illasration of the effect of parameters on the price of corporate bond and the credit spread.

{\bf Acknowledgment} Authors thank anonymous arXiv moderators for strict note which helps to make this version better and more complete.


\end{document}